\documentstyle[11pt,amssymb,amsfonts,amsthm,amssymb,amscd,amstext,epsfig]{article}

\def\ben{\begin{equation}}
\def\een{\end{equation}}

  \let\g=\gamma  
    
 \let\m=\mu \let\n=\nu

\let\C=\Chi

 \def\bd{\begin{document}} \def\ed{\end{document}}
\def\ds{\documentstyle} \let\fr=\frac \let\bl=\bigl \let\br=\bigr
\let\Br=\Bigr \let\Bl=\Bigl
\let\bm=\bibitem
\let\na=\nabla
\let\pa=\partial \let\ov=\overline
\newcommand{\be}{\begin{equation}}
\newcommand{\ee}{\end{equation}}
\def\ba{\begin{array}}
\def\ea{\end{array}}
\def\ft#1#2{{\textstyle{{\scriptstyle #1}\over {\scriptstyle #2}}}}
\def\fft#1#2{{#1 \over #2}}
\def\del{\partial}
\def\vp{\varphi}
\def\sst#1{{\scriptscriptstyle #1}}
\def\oneone{\rlap 1\mkern4mu{\rm l}}
\def\td{\tilde}
\def\wtd{\widetilde}
\def\ie{\rm i.e.\ }
\def\dalemb#1#2{{\vbox{\hrule height .#2pt
        \hbox{\vrule width.#2pt height#1pt \kern#1pt
                \vrule width.#2pt}
        \hrule height.#2pt}}}
\def\square{\mathord{\dalemb{6.8}{7}\hbox{\hskip1pt}}}
\newcommand{\ho}[1]{$\, ^{#1}$}
\newcommand{\hoch}[1]{$\, ^{#1}$}
\newcommand{\bea}{\begin{eqnarray}}
\newcommand{\eea}{\end{eqnarray}}
\newcommand{\ra}{\rightarrow}
\newcommand{\lra}{\longrightarrow}
\newcommand{\Lra}{\Leftrightarrow}
\newcommand{\ap}{\alpha^\prime}
\newcommand{\bp}{\tilde \beta^\prime}
\newcommand{\tr}{{\rm tr} }
\newcommand{\Tr}{{\rm Tr} }
\def\0{{\sst{(0)}}}
\def\1{{\sst{(1)}}}
\def\2{{\sst{(2)}}}
\def\3{{\sst{(3)}}}
\def\4{{\sst{(4)}}}
\def\5{{\sst{(5)}}}
\def\6{{\sst{(6)}}}
\def\7{{\sst{(7)}}}
\def\8{{\sst{(8)}}}
\def\n{{\sst{(n)}}}
\def\cA{{{\cal A}}}
\def\cF{{{\cal F}}}
\def\tV{\widetilde V}
\def\tW{\widetilde W}
\def\tH{\widetilde H}
\def\tE{\widetilde E}
\def\tF{\widetilde F}
\def\tA{\widetilde A}
\def\im{{{\rm i}}}
\def\tY{{{\wtd Y}}}
\def\ep{{\epsilon}}
\def\vep{{\varepsilon}}
\def\R{\rlap{\rm I}\mkern3mu{\rm R}}
\def\bD{{{\bar D}}}

\def\R{\rlap{\rm I}\mkern3mu{\rm R}}
\def\bD{{{\bar D}}}
\def\R{{{\Bbb R}}}
\def\C{{{\Bbb C}}}
\def\H{{{\Bbb H}}}
\def\CP{{{\Bbb C}{\Bbb P}}}
\def\RP{{{\Bbb R}{\Bbb P}}}
\def\Z{{{\Bbb Z}}}
\def\bA{{{\Bbb A}}}
\def\bB{{{\Bbb B}}}
\def\bC{{{\Bbb C}}}
\def\bR{{{\Bbb R}}}
\def\bD{{{\Bbb D}}}
\def\bE{{{\Bbb E}}}
\def\bZ{{{\Bbb Z}}}
\def\bQ{{{\Bbb Q}}}
\def\Re{{{\frak{Re}}}}
\def\Im{{{\frak{Im}}}}
\def\cosec{{\,\hbox{cosec}\,}}
\def\Gm{{\Gamma_{\!\! -}}}
\def\Gp{{\Gamma_{\!\! +}}}

\def\cosech{{\hbox{cosech}}}

\def\etcyc{{\hbox{and cyclic}}}

\newtheorem{thm}{Theorem}
\newtheorem{lem}[thm]{Lemma}
\newtheorem{cor}[thm]{Corollary}

\begin{document}
\begin{flushright}
\hfill{DAMTP-2003-12} \\
{hep-th/0302072}
\end{flushright}

\begin{center}
\vspace{1cm} { \Large {\bf Compactification, topology change and
surgery theory}}

\vspace{1.5cm}

Sean A. Hartnoll

\vspace{0.3cm}

s.a.hartnoll@damtp.cam.ac.uk

\vspace{0.8cm}

{\it DAMTP, Centre for Mathematical Sciences,
 Cambridge University\\ Wilberforce Road, Cambridge CB3 OWA, UK}

\vspace{2cm}

\end{center}

\begin{abstract}
We study the process of compactification as a topology change. It
is shown how the mediating spacetime topology, or cobordism, may
be simplified through surgery. Within the causal Lorentzian
approach to quantum gravity, it is shown that any topology change
in dimensions $\geq 5$ may be achieved via a causally continuous
cobordism. This extends the known result for $4$ dimensions.
Therefore, there is no selection rule for compactification at the
level of causal continuity. Theorems from surgery theory and
handle theory are seen to be very relevant for understanding
topology change in higher dimensions. Compactification via
parallelisable cobordisms is particularly amenable to study with
these tools.
\end{abstract}

\pagebreak
\setcounter{page}{1}

\tableofcontents
\addtocontents{toc}{\protect\setcounter{tocdepth}{3}}

\section{Introduction}

The requirement of extra spacetime dimensions is a persistent
theme in theories of unification and of quantum gravity. For consistency
with experience, the theory must contain a mechanism that would
allow only four dimensions to have been observed thus far. Compactification
of extra dimensions is thought to be one such mechanism. In this paper we study the topology
involved in the process of compactification \cite{Tipler:qu}.

By a compactified universe, we mean a universe with spacelike
topology $S^{d-1}\times A$. The sphere $S^{d-1}$ is thought of as
the large, observed, space directions and the internal space $A$,
of dimension $n-d$, is small and unobserved. Note $n$ is the total
dimension of spacetime and $d$ is the observed dimension of
spacetime. A universe with spacelike topology $S^{n-1}$ is `not
compactified' and all dimensions are large.

It seems reasonable to think that, in some regime at least, quantum gravity may
be conceived as a path integral over spacetimes \cite{Hawking:jz}. One considers all topologies
joining the given initial and final spacelike hypersurfaces. Or, in the context of
creation of the universe from nothing \cite{Vilenkin:rn}, one considers all topologies with
a given single boundary hypersurface.

The process of compactification is one in which the initial
spacelike topology is $M=S^{n-1}$ and the final topology is
$M^{\prime} = S^{d-1}\times A$. An immediate consistency
requirement is that there exists a manifold $X$ with boundaries
$M$ and $M^{\prime}$. Consider filling in the $S^{n-1}$, to give
the disc $D^n$. We now see that the question is the same as
requiring the existence of a manifold $X$ with unique boundary
$\pa X = M^{\prime}$, describing the creation from nothing of a
compactified universe. Such issues are the concern of cobordism
theory, which is introduced briefly below.

The manifold $X = D^d\times A$, where $D^d$ is the $d$-dimensional disc, clearly
provides a manifold with boundary $S^{d-1}\times A$, as required. No internal
manifold $A$ is ruled out at this level. However,
for various reasons, one would like a more systematic understanding of compactification.

The most pressing concern arises in a causal Lorentzian approach
to quantum gravity. Here, the intermediate manifold, or cobordism,
$X$, is considered together with an almost Lorentzian metric that
gives a well-defined causal structure on $X$. Well-known theorems
forbid causal topology change with a fully Lorentzian metric
\cite{geroch,Tipler:qu}. Almost Lorentzian metrics evade these
theorems by allowing the metric to be degenerate at certain points
\cite{Horowitz:qb,sorkin1,Vilenkin:rn}. Whilst such almost
Lorentzian metrics always exist, it is conjectured
\cite{Dowker:1997hj} that the structure only contributes to the
path integral if it is causally continuous, as described below. It
has been shown that the requirement of a causally continuous
almost Lorentzian metric is equivalent to certain topological
conditions on $X$ \cite{sorkin2,Dowker:1999wu}. Thus, in order for
a given compactification to be possible, one needs to prove the
existence of a manifold $X$ satisfying the required conditions.
Generically, $X = D^d\times A$ will not satisfy these conditions.

Even outside the causal Lorentzian approach, one would like to know what kind of topologies
are possible for $X$. Is there a `simplest' allowed topology? Adding dynamics to the
system at a later stage may impose further topological restrictions. We will obtain
cobordisms $X^{\prime}$, with different topology to $X$, by {\bf surgery} on $X$. The
causal structure of $X^{\prime}$ is then studied using {\bf handle decomposition}.

Surgery theory and handle decomposition have been used previously
in discussions of topology change in the physics literature, see for example
\cite{Dowker:1997kc,Dowker:1997hj,Ionicioiu:1997hi,Alty:1994xs,Konstantinov:jn}.
Previous notes on higher dimensions and topology change are \cite{Tipler:qu,Ionicioiu:1997zy}.
However, many of the powerful theorems in these areas of mathematics
were not used. We will see that some of these theorems are very useful for studying
higher dimensional topology change.

One important result of this work is that any compactification
$S^{d-1}\times A$ may be obtained via a manifold $X$ that admits a
causally continuous Lorentzian metric, if the total dimension of
spacetime $n \geq 5$. Thus, no compactification is ruled out at
this level in causal Lorentzian quantum gravity. Indeed the result
is more general, and {\it any} topology change in these
dimensions, between connected initial and final topologies, that
is possible on cobordism grounds may be achieved in a causally
continuous way. This result was already known for $n=4$ spacetime
dimensions \cite{Dowker:1997hj}.

Another important point is that when compactifying on group
manifolds, and in general via parallelisable cobordisms, we will
see that it is possible to systematically rearrange the cobordisms
for the process of compactification.

Section 2 reviews mathematical results from relative homology and
homotopy theory, surgery theory,
handle decomposition and Morse theory. Section 3 shows how the topology of the
cobordism for a generic compactification may be rearranged such that the resulting
cobordism admits a causally continuous almost Lorentzian metric. Section 4 considers
the case of compactification via parallelisable cobordisms, where significantly further rearrangement
is possible. Section 5 contains remarks on when all cobordisms can be obtained via surgery
on the initial cobordism $X$. Section 6 is a concluding discussion. The appendix
contains an argument for why surgery is a natural operation to consider on a manifold.

\section{Mathematical preliminaries}

All manifolds discussed are assumed to be smooth, connected
and compact.

\subsection{Results from relative homology and homotopy theory}

Let $A$ be a subspace of the topological space $X$, and $S_k(X)$
the free abelian group of k-chains in $X$. Let $\pa_k : S_k(X) \to S_{k-1}(X)$
be the boundary operator. The group of {\bf relative k-cycles} mod $A$ is
\ben
Z_k(X,A) = \{ \g \in S_k(X) : \pa_k \g \in S_{k-1}(A) \} .
\een
The group of {\bf relative k-boundaries} mod $A$ is
\ben
B_k(X,A) = \{ \g \in S_k(X) : \g - \g^{\prime} \in B_k(X), \mbox{ some } \g^{\prime} \in S_k(A) \} ,
\een
where $B_k(X)$ is the usual group of k-boundaries. The {\bf relative homology} groups
are now defined as
\ben
H_k(X,A) = Z_k(X,A) / B_k(X,A) .
\een
We will be mostly interested in the case when $A = \pa X$, the boundary of $X$.
The two following theorems from relative homology theory will be used. For
extended discussions see e.g. \cite{rotman,maunder}.

\begin{thm} Given a subspace $A$ of $X$, there exists an exact sequence

$\begin{CD}
\cdots  @>>> H_k(A) @>>> H_k(X) @>>> H_k(X,A) @>>> H_{k-1}(A) @>>> \cdots .
\end{CD}$\label{thm:exact}
\end{thm}

\begin{thm} {\bf (Lefschetz duality)} Given an oriented manifold $X$ of dimension $n$ and
boundary $\pa X$ then one has the isomorphisms $H^k(X) \cong
H_{n-k}(X, \pa X)$ and $H^k(X, \pa X) \cong H_{n-k}(X)$ for all $k$.
\label{thm:lef}
\end{thm}

It can be useful to combine Lefschetz duality with the isomorphism of vector spaces $H^k(X) \cong H_k(X)$.
This isomorphism only holds with
coefficients in a field, it does not hold when there is
torsion. However, the following theorem shows the precise effect of
torsion \cite{rotman}:

\begin{thm} {\bf (Universal coefficient for cohomology)} The sequence, with
coefficients of the homology and cohomology groups shown explicitly,

$\begin{CD}
0  @>>> \text{{\rm Ext}}^1_{\bZ}(H_{k-1}(X,\bZ),\bZ) @>>> H^k(X,\bZ)
@>>> \text{{\rm Hom}}_{\bZ}(H_k(X,\bZ),\bZ) @>>> 0 ,
\end{CD}$

\noindent is exact (and splits).
\label{thm:coeff}
\end{thm}

All we need to know about the first term in the sequence,
$\text{Ext}^1_{\bZ}(H_{k-1}(X,\bZ),\bZ)$, is that it gives the
torsion of $H_{k-1}(X,\bZ)$. For the full definition, see for
example \cite{rotman}. The last term gives the non-torsion part of
$H_k(X,\bZ)$. Thus, the theorem is saying that a copy of $\bZ$ in
$H_k(X,\bZ)$ contributes $\bZ$ to $H^k(X,\bZ)$ and a copy of
$\bZ_p$ in $H_{k-1}(X,\bZ)$ contributes $\bZ_p$ to $H^k(X,\bZ)$.

One also has relative homotopy spaces, $\pi_k(X,A)$. Most important
will be $\pi_1(X,A)$, which is not a group and is given by the
homotopy classes relative to $A$ of paths in $X$ that have one endpoint at a basepoint
$x_0 \in A$ and the other endpoint in $A$. The higher relative
homotopy spaces are groups \cite{rotman}. The following two theorems
are important \cite{rotman,maunder}:

\begin{thm} Given a subspace $A$ of $X$, there exists an exact sequence

$\begin{CD}
\cdots  @>>> \pi_k(A) @>>> \pi_k(X) @>>> \pi_k(X,A) @>>> \pi_{k-1}(A)
@>>> \cdots @>>> \pi_0(X) .
\end{CD}$\label{thm:exacthtp}
\end{thm}

\begin{thm} {\bf (Hurewicz)} If $\pi_k(X,A)=0$ for $1\leq k \leq s-1$,
then $H_k(X,A)=0$ for $1\leq k \leq s-1$. If, further,
$\pi_1(X)=\pi_1(A)=0$, then $\pi_s(X,A)\cong H_s(X,A)$.
\label{thm:hur}
\end{thm}

\subsection{Surgery theory and cobordism}

An oriented manifold $X$ is an {\bf oriented cobordism} between the oriented manifolds $M$
and $M^{\prime}$ if $\pa X$, with the induced orientation, is diffeomorphic to the disjoint union
of $M$ and $-M^{\prime}$. Here, $-$ denotes orientation reversal. Cobordism defines
an equivalence relation on the space of oriented manifolds. We will generally be interested in
the case $M=\emptyset$. Thus the manifold $X$ has connected boundary
$\pa X = M^{\prime}$, neglecting the change in orientation. There is a
simple criterion for when two manifolds without boundary are
oriented-cobordant \cite{milnor2,milnor3}:

\begin{thm} Let $M,M^{\prime}$ be manifolds without boundary. Then
$M$ and $M^{\prime}$ are oriented-cobordant if and only if they have
the same Pontrjagin and Stiefel-Whitney numbers.

\label{thm:cobor}
\end{thm}

Given a cobordism $X$, it is possible to obtain a different cobordism $X^{\prime}$, with
$\pa X = \pa X^{\prime}$, through surgery on $X$. Suppose $X$ is $n$ dimensional. Intuitively,
surgery, also known as spherical modification, should be thought of as removing an embedded
sphere of dimension $k$ and replacing it with an embedded sphere of dimension $n-k-1$.
A more precise description is as follows \cite{milnor1,wallace1,hcobordism}.

Start with an embedding of $\phi : S^k\times D^{n-k} \to X$. The boundary of the embedding is
$S^k\times S^{n-k-1}$, which is also the boundary of $D^{k+1}\times S^{n-k-1}$. We may thus
remove the interior of the embedding and replace it with the interior of $D^{k+1}\times S^{n-k-1}$.
The result is the manifold
\ben
X^{\prime} = \left(X - \phi (S^k\times 0) \right) + \left(D^{k+1}\times S^{n-k-1} \right) ,
\een
where $-$ denotes removal and $+$ denotes an identification of $\phi(u,\theta v)$ with $(\theta u, v)$ for each $u\in S^k, v
\in S^{n-k-1}$ and $0 < \theta \leq 1$. This will be called a {\bf type $(k,n-k-1)$ surgery}.
The process is illustrated in Figure 1.
\begin{figure}[h]
\begin{center}
\vspace{4mm}
\epsfig{file=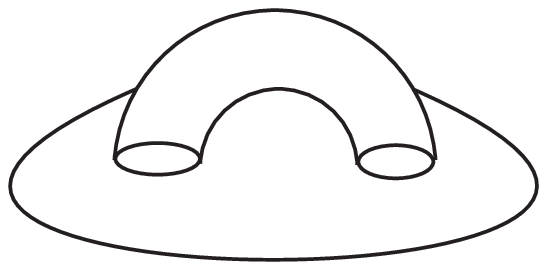,width=2cm} $\to $
\epsfig{file=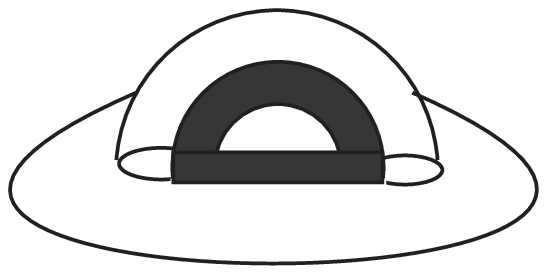,width=2cm} $\to $
\epsfig{file=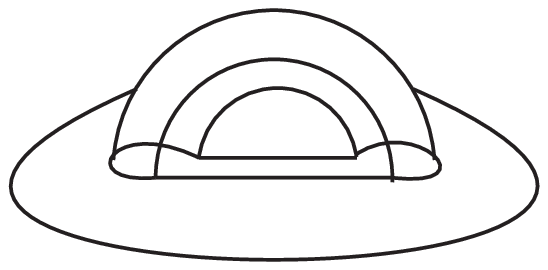,width=2cm} $\to $
\epsfig{file=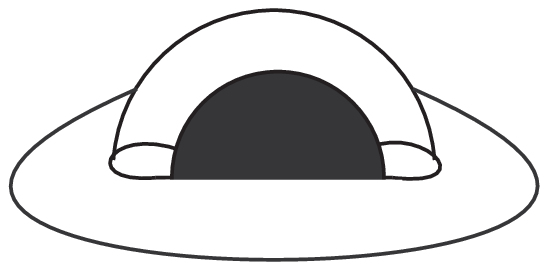,width=2cm} $\to $
\epsfig{file=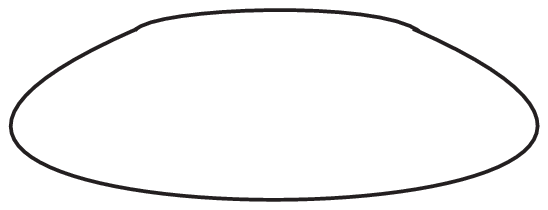,width=2cm}
\end{center}
\noindent {\bf Figure 1:} Surgery between $X$ and $X^{\prime}$, both with boundary $S^1$.
An $S^1\times D^1$ is removed and replaced with a $D^2\times S^0$. The change
in topology is evident.
\end{figure}

Note that we are using surgery to modify the cobordism itself. This should not be confused with
the use of surgery to construct cobordisms by modifying manifolds without boundaries.
By performing surgeries below, we will find cobordisms for quantum compactification with desirable
topological properties.

\subsection{Handle decomposition, Morse theory and causal continuity}

Once we have obtained an interesting cobordism $X$, it will be useful to consider its
handle decomposition \cite{rourke,hcobordism,Dowker:1997kc}. A {\bf handle of index $k$}
on an $n$ dimensional manifold, $X$, is an $n$-disc $D^n$ such that $X \cap D^n \subset \pa X$,
and there is a homeomorphism $h: D^k \times D^{n-k} \to D^n$, such that
$h(S^{k-1}\times D^{n-k}) = X \cap D^n$. Where $\pa D^k = S^{k-1}$. Two simple examples are
shown in Figure 2. Adding a handle is closely related to performing a surgery, as we shall see below.
\begin{figure}[h]
\begin{center}
\vspace{4mm}
\epsfig{file=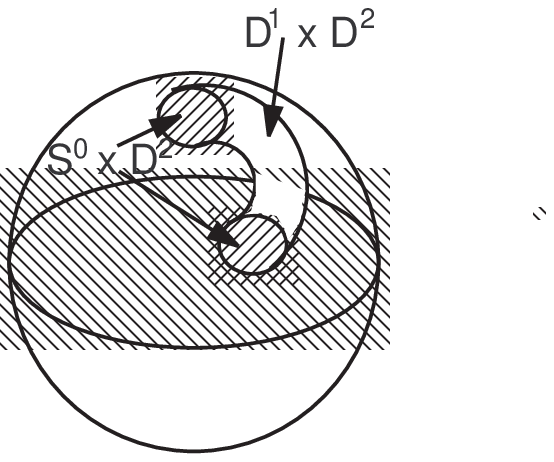,width=3cm} $\to $
\epsfig{file=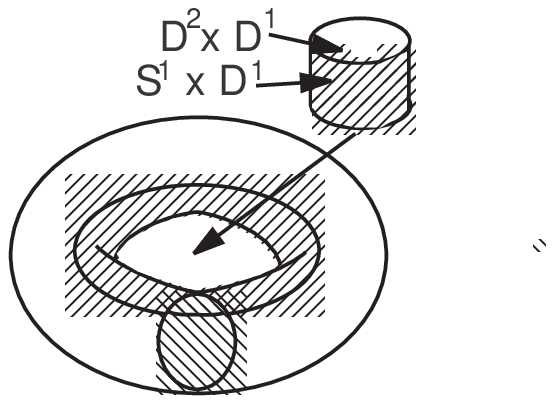,width=3cm} $\to $
\epsfig{file=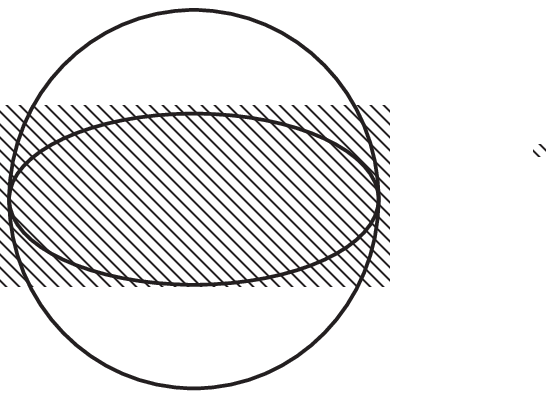,width=3cm}
\end{center}
\noindent {\bf Figure 2:} Adding a 1-handle to $D^2$ to obtain a solid torus. Adding a 2-handle
to a solid torus to reobtain the ball.
\end{figure}

A {\bf handle decomposition} of a cobordism, $X$ from $M$ to $M^{\prime}$, is a presentation
\ben
X = C_0 \cup H_1 \cup \cdots \cup H_t ,
\label{eq:handle}
\een
where $C_0 = M \times [0,1]$ and $H_k$ is a handle on the cobordism
\ben
X_{k-1} = C_0 \cup \{ \cup H_l \mid l \leq k-1 \} .
\een
This gives a procedure for constructing $X$ from the trivial cobordism. If $\pa X$
has a single connected component, $M^{\prime}$, then one may start from the disc $C_0 = D^n$.
The handle decomposition of $X$ is not unique. For example, Figure 2 shows
a handle decomposition of a 2-disc as a 2-disc with a 1-handle and a 2-handle added.
Handle decompositions are generic by the following crucial result \cite{rourke}:

\begin{thm} Every cobordism admits a handle decomposition .
\label{thm:han}
\end{thm}

Handle decomposition is also closely related to Morse theory \cite{hcobordism,morse}. Morse theory will be used to define
an almost Lorentzian metric on the cobordism with certain causal properties. A function $f : X \to \bR$ has
a {\bf critical point} at $p\in X$ if $\pa_i f(p) = 0$. The critical point is non-degenerate if
$\mbox{det} \left[ \pa_i \pa_j f(p) \right] \neq 0$. A {\bf Morse function}
on a cobordism $X$ is a function $f : X \to \bR$ that is constant on each connected component of $\pa X$
and whose critical points are in the interior of $X$ and non-degenerate. Every cobordism admits a Morse function,
a result closely related to Theorem \ref{thm:han}.

The {\bf index} of a non-degenerate critical point $p$ is the number of negative eigenvalues
of the Hessian $\pa_i \pa_j f (p)$. The number of critical points with index $k$ will be denoted
$\m_k(f)$. The following result is important (this is theorem 3.12 of \cite{hcobordism}
translated into the language of handles):

\begin{thm} Given a handle decomposition of the cobordism $X$, then $X$ admits a Morse function with
exactly one critical point of index $k$ for each $k$-handle in the decomposition.
\label{thm:handles}
\end{thm}

The power of this result is that it gives us an equality for the
number of critical points of a Morse function. This should be
contrasted with the well-known weak Morse inequalities
\cite{morse} $b_k \leq \m_k(f)$, where $b_k$ are the Betti numbers
of the manifold, $X$.

Given a Morse function $f$ on $X$ and a Riemannian metric $G$ on
$X$, which always exists, one may then construct an almost
Lorentzian metric \cite{sorkin1} \ben g_{\m\nu} = G^{\rho
\sigma}\pa_\rho f \pa_\sigma f G_{\m \nu} - \zeta \pa_\m f \pa_\nu
f , \een where $\zeta > 1$ is a real number. This metric is
Lorentzian everywhere except at the critical points and has a
well-defined causal structure because $f$ acts as a time function.
The timelike direction is $G^{\m \nu}\pa_\nu f$. This almost
Lorentzian metric is said to define a Morse spacetime.

The final idea we need is that of {\bf causal continuity}
\cite{hawking}. Intuitively, a spacetime is causally discontinuous
if the volume of the causal past or future of some point changes
discontinuously under a continuous change in the point. A causally
continuous spacetime may be characterised by the condition \ben x
\in \overline{I^-(y)} \, \Leftrightarrow \, y \in
\overline{I^+(x)}, \quad\quad  \forall x,y \in X ,
\label{eq:cause}\een Where $\overline{I^{\pm}(p)}$ denotes the
closure of the chronological past and future of $p$. Recall that
the chronological future of a point consists of all other points
that may be reached via timelike curves. The closure in equation
(\ref{eq:cause}) is what makes the definition work.

It is conjectured \cite{Dowker:1997hj} that causally discontinuous
spacetimes do not contribute to the Lorentzian sum over histories.
It was further conjectured \cite{Dowker:1997hj} that causal
continuity should be associated with critical points of index 1
and $n-1$ of Morse functions. It was later proven
\cite{Dowker:1999wu,sorkin2} that:

\pagebreak

\begin{thm} If all Morse functions on a cobordism $X$ contain critical points of index 1
or $(n-1)$, then the cobordism supports only causally discontinuous Morse spacetimes.
Conversely, if $X$ admits a Morse function with no critical points of index 1 or $(n-1)$, then it does
support causally continuous Morse spacetimes.
\label{thm:causal}
\end{thm}

Thus, putting the above results together, in the context of causal
Lorentzian quantum gravity, there is a selection rule for topology
change. Topology change requires a cobordism with a handle
decomposition with no 1-handles or $(n-1)$-handles. We will see
that this requirement is trivial in dimensions $n \geq 5$. It is
already known in the physics literature to be trivial in dimension
$n=4$ due to the Lickorish-Wallace theorem for three-manifolds
\cite{Dowker:1997hj}.

\section{Quantum compactification: generic case}

In this section we will see that there exists a simply connected cobordism for the
quantum creation of any compactified universe, $S^{d-1}\times A$, from nothing.
We will then see that this cobordism admits a causally continuous
almost Lorentzian metric.

An obvious cobordism for the creation of the compactified universe
is $X = D^{d}\times A$, with boundary $S^{d-1}\times A$. From this starting point we will
then perform surgery, if necessary, to obtain a simply connected cobordism $X^{\prime}$.
Given $X^{\prime}$ one can then consider a handle decomposition of the cobordism
and show that it admits a Morse function with no critical points of index $1$ or $n-1$.
The process is summarised as follows:

\begin{center} $\begin{CD}
D^d\times A @>\text{surgery}>> X^{\prime},\, \pi_1(X^{\prime})=0
@>\text{decomp.}>> f : X^{\prime} \to \bR,\, \m_1(f) = \m_{n-1}(f)
= 0 .
\end{CD}$
\end{center}
\vspace{0.2cm}

We will work in a more general framework, in which $X$ is a cobordism
between $M$ and $M^{\prime}$. In the compactification case
$M=\emptyset$ and $M^{\prime}=S^{d-1}\times A$.

\subsection{Eliminating the fundamental group}

The possibility of eliminating the fundamental group through surgery is contained in the following
theorem due to Milnor \cite{milnor1} and Wallace \cite{wallace1}:

\begin{thm} An oriented compact manifold, with or without boundary and of dimension $n\geq 4$,
can be made simply connected through a finite sequence of
surgeries of type $(1,n-2)$.\label{thm:simply}
\end{thm}

The result of the surgeries, $X^{\prime}$, thus satisfies
$H_1(X^{\prime})=\pi_1(X^{\prime}) = 0$.

In fact a little more is true. Suppose $X$, hence $X^{\prime}$, is a
cobordism between $M$ and $M^{\prime}$. Then we can now easily see that
$\pi_1(X^{\prime},M) = \pi_1(X^{\prime},M^{\prime})=0$.
Consider the end of the exact sequence of Theorem \ref{thm:exacthtp},
given that $\pi_1(X^{\prime})=0$:
\ben
\begin{CD}
0 @>>> \pi_1(X^{\prime},M^{\prime}) @>>> \pi_0(M^{\prime}) @>>>
\pi_0(X^{\prime}) . \label{eq:con}
\end{CD}
\een
However, $\pi_0(M^{\prime}) \cong \pi_0(X^{\prime})$, because
$M^{\prime}$ and $X^{\prime}$ are connected and the inclusion map
induces the isomorphism. Exactness then implies that
$\pi_1(X^{\prime},M^{\prime}) = 0$. This result is also
straightforward to see directly. If $M^{\prime}=\emptyset$ then we
get the same result because $\pi_1(X^{\prime},M^{\prime}) =
\pi_1(X^{\prime}) = 0$. Clearly the same argument may be used for
$M$ instead of $M^{\prime}$. The following subsection will further
consider the cases when $X^{\prime}$, $M$ or $M^{\prime}$ are not
connected.

In the appendix, a motivation is given for considering surgery as
a `minimal' transformation on manifolds that may eliminate
homotopy groups.

\subsection{A causally continuous Morse metric for compactification}

We have obtained a cobordism for compactification, $X^{\prime}$,
with
$\pi_1(X^{\prime})=\pi_1(X^{\prime},M)=\pi_1(X^{\prime},M^{\prime})=0$.
This now allows us to find a particularly amenable handle
decomposition of $X^{\prime}$. The important statement in this
subsection is Theorem \ref{thm:1han} below, but some preliminary
remarks concerning connectivity, 0-handles and $n$-handles are
needed.

First note that without loss of generality, we may assume that the
initial cobordism $X$, and hence $X^{\prime}$, is connected.
Suppose $X$ were not connected. Then we could make it connected by
use of type $(0,n-1)$ surgeries. Concretely, in order to connect
two components, remove a small disc from each, $S^0\times D^n$,
and then join the components with $D^1\times S^{n-1}$. Once $X$ is
connected, connectivity of $X^{\prime}$ follows because
connectivity is not affected by the type $(1,n-2)$ surgeries of
Theorem \ref{thm:simply}. In fact, we don't strictly need to do
this, we could work with a disconnected cobordism by considering
each connected component to give an independent cobordism, and
then applying the argument below to each component separately. We
may now use the following result \cite{rourke}:

\begin{lem} If $X^{\prime}$ is a connected cobordism from $M$ to
$M^{\prime}$, then $X^{\prime}$ admits a handle decomposition with
no 0-handles if $M \neq \emptyset$ and no $n$-handles if
$M^{\prime} \neq \emptyset$.\label{thm:0handles}
\end{lem}

The case $M = \emptyset$ or $M^{\prime} = \emptyset$ do not pose a
problem for our purposes. In, say, the case $M = \emptyset$,
remove a small disc $D^n$ from $X^{\prime}$. Now view $X^{\prime}$
as broken into two sequential cobordisms; the first, $D^n$, from
$\emptyset$ to $S^{n-1}$, and the second, $X^{\prime} - D^n$, from
$S^{n-1}$ to $M^{\prime}$. The first of these cobordisms is
trivial, made up of a 0-handle and nothing else. The second
cobordism now fulfills the criteria of Lemma \ref{thm:0handles}.
In the case where originally we had $X^{\prime}$ with
$M=\emptyset$, we shall instead regard $X^{\prime}$ as being the
second cobordism, with the $D^n$ excised, and therefore
$M=S^{n-1}$. We will then simply add the $D^n$ back in after
applying the following theorem \cite{rourke,hcobordism}:

\begin{thm} Suppose $X^{\prime}$ is a connected cobordism from $M$ to $M^{\prime}$
with a handle decomposition with no 0-handles and $i_k$
$k$-handles for $k>0$. Suppose $\pi_1(X^{\prime},M)=0$ and $n \geq
5$. Then $X^{\prime}$ admits a different handle decomposition with
$i_k$ handles for $k \neq 1,3$, no 1-handles and $(i_1+i_3)$
3-handles.
\label{thm:1han}
\end{thm}

Thus we may always find a handle decomposition for $X^{\prime}$
which has no 1-handles. Adding back in an excised $D^n$, if
necessary, adds only a 0-handle to the decomposition, and no
1-handles. Further, by considering the `dual handle decomposition'
\cite{rourke}, which views the cobordism in the opposite
direction, from $M^{\prime}$ to $M$, Theorem \ref{thm:1han} may
further be used to remove the $(n-1)$-handles, because
$\pi_1(X^{\prime},M^{\prime})=0$. By Theorems \ref{thm:handles}
and \ref{thm:causal} we then see the cobordism $X^{\prime}$ admits
a causally continuous almost Lorentzian metric. No
compactification is ruled out due to causal discontinuities for
dimensions $n\geq 5$.

More generally, we see that no topology change in higher
dimensions, with connected $M$ and $M^{\prime}$, is ruled out
through considerations of causal continuity. Some topology change
will be ruled out on the grounds of Theorem \ref{thm:cobor}. By
constructing the initial cobordism $X$ explicitly above, we saw
that this is never a problem for compactification. The condition
of $n\geq 5$ arises from uses of the Whitney Lemma \cite{rourke},
which is used to rearrange handles so that the intersections of
handles may be placed in a canonical form.

In the above we have assumed that the initial and final
topologies, $M$ and $M^{\prime}$, are connected. Consider the case
that $M^{\prime}$ has various disconnected components. The key
step above that no longer holds is that following equation
(\ref{eq:con}); if $X^{\prime}$ is connected, then
$\pi_0(M^{\prime})$ and $\pi_0(X^{\prime})$ are no longer
isomorphic. It follows from (\ref{eq:con}) that
$\pi_1(X^{\prime},M^{\prime})$ is nonzero, contrary to the
requirements of our argument for eliminating $(n-1)$-handles.

In order to solve this problem, we would need $X^{\prime}$ to be
disconnected, with one component containing each component of
$M^{\prime}$. This would restore the desired isomorphism
$\pi_0(M^{\prime}) \cong \pi_0(X^{\prime})$. We
could now think of each connected component of $X^{\prime}$ as a
cobordism in itself and apply Theorem \ref{thm:1han} to each
component separately. Unfortunately, whilst we have seen above
that any two cobordant manifolds are cobordant via a connected
cobordism, it is not necessarily true that there will exist a
disconnected cobordism with the properties that we have just
required. In particular, suppose $M=\emptyset$ and $M^{\prime} = N
\sqcup N^{\prime}$ has two components. Suppose further that
neither $N$ nor $N^{\prime}$ are cobordant to the empty set, that
is, they do not bound. Then there does not exist a cobordism with
two components with boundaries $N$ and $N^{\prime}$ because, by
assumption, $N$ and $N^{\prime}$ do not bound. For example, let $N
= - N^{\prime} = {\mathbb{CP}}^2$. Here $-$ denotes orientation
reversal. If we wanted to work only with spin manifolds, then we
should use spin cobordism rather than oriented cobordism. In this
case an example, in eight spatial dimensions, would be $N = -
N^{\prime} = {\mathbb{HP}}^2$.

To summarise, our argument will not work in circumstances such as
when $M=\emptyset$ and when $M^{\prime}$ is disconnected and with
two or more components that do not individually bound. Thus,
requiring causal continuity might forbid the simultaneous creation
of multiple universes with topologies that do not bound.

\section{Compactification via parallelisable cobordisms}

In this section we see that when the cobordism, $X$, is a parallelisable manifold, then the
topology for quantum compactification may be rearranged considerably
further than in the generic case. Recall that Lie groups are parallelisable.
Note that this class of spacetimes includes torus
compactifications. Weaker conditions than parallelisability are
possible \cite{milnor1}, but will not be studied here.

\subsection{Eliminating higher homotopy groups}

For a generic manifold, eliminating the fundamental group was the only possibility. This is
because in order to perform the surgeries that eliminate homotopy
groups, $\pi_k(X)$, one needs to
use embedded spheres with trivial normal bundle, in order to embed $S^k\times D^{n-k}$.
For any orientable manifold, an embedded $S^1$ will have trivial normal bundle,
but this will not be true for higher dimensional embedded spheres.

However, if the cobordism has additional properties, then it will be possible to find spheres
representing homotopy classes that have trivial normal bundles. In this case one
can use surgery to eliminate the homotopy class by collapsing the sphere.

Following \cite{milnor1,kervaire}, define a manifold, $X$, to be
{\bf S-parallelisable} if the sum $T(X)\oplus L$ is a trivial
bundle. Here $T(X)$ is the tangent bundle and $L$ is a trivial
line bundle over $X$. One now has the following lemmas
\cite{kervaire}:

\begin{lem} If $X$ is a submanifold of $S^{n+m}$, $n<m$, then $X$ is S-parallelisable
if and only if its normal bundle is trivial.
\end{lem}

\begin{lem} A connected manifold with non-vacuous boundary is S-parallelisable if and
only if it is parallelisable.
\end{lem}

However, it is an elementary fact that a Lie group is a
parallelisable manifold. Further, $D^d$ is also clearly
parallelisable and therefore one has:

\begin{cor} The cobordism $X=D^d\times A$ is S-parallelisable, for any Lie group $A$.
\end{cor}

An example of a parallelisable manifold that is not a Lie group is $S^7$.
The usefulness of S-parallelisability is that it allows us to eliminate many homotopy groups
\cite{milnor1,kervaire,wallace1}:

\begin{thm} Let $X$ be compact, connected, and S-parallelisable, of dimension $n \geq 2m$.
Then, by a sequence of surgeries, one can obtain an
S-parallelisable manifold, $X^{\prime}$, with $\pi_k(X^{\prime}) =
0$ for all $k \leq m-1$. \label{thm:spar}
\end{thm}

A consequence of this theorem is that the homology of $X^{\prime}$ is largely determined by that
of the boundary $\pa X^{\prime}=\pa X$ using Theorems
\ref{thm:lef} and \ref{thm:coeff}. Theorem \ref{thm:hur}
implies that $\pi_k(X^{\prime}) = 0, \, 0<k\leq m-1 \Rightarrow H_k(X^{\prime}) = 0, \, 0 <k\leq m-1$.
One has, for even dimensions $n=2m$,
\bea
\label{eq:homol}
H_0(X^{\prime}) & = & \bZ , \nonumber \\
H_k(X^{\prime}) & = & 0 \quad \quad 1 \leq k \leq m - 1 ,\nonumber \\
H_m(X^{\prime}) & & \mbox{ not determined from boundary } ,\nonumber \\
H_{m+k}(X^{\prime}) & = & H_{m+k}(\pa X) \quad \quad 1 \leq k \leq m - 2 ,\nonumber \\
H_{2m-1}(X^{\prime}) & = & H_{2m}(X^{\prime}) = 0 . \eea These
results follow from Theorem \ref{thm:spar} as an elementary
exercise using Theorems \ref{thm:exact}, \ref{thm:lef} and
\ref{thm:coeff}. The final line of (\ref{eq:homol}) requires $\pa
X^{\prime}$ to be connected. A similar list is possible in odd
dimensions. The only difference is that if $n=2m+1$, then both
$H_m(X^{\prime})$ and $H_{m+1}(X^{\prime})$ are not determined by
the boundary.

This elimination of homology groups is all we shall use below.
Note however, that the vanishing of homology groups has
implications for handle decompositions \cite{rourke}:

\begin{thm} Let $X$ be a cobordism from $M$ to $M^{\prime}$ with no handles of index $<s$ and $i_t$
$t$-handles for $t \geq s$. Then, if $M$ is simply connected and $2 \leq s \leq n-4$,
$n \geq 6$ and $H_s(X,M) = 0$, we can find a new handle decomposition with the
same number of $t$-handles for $t\neq s,s+1$, with no $s$-handles and with $i_{s+1}-i_s$
$(s+1)$-handles.
\end{thm}

This theorem implies that for an S-parallelisable manifold one may
find a handle decomposition on $X^{\prime}$ with no $k$-handles
for $k \leq m-1$. Note that here $M=\emptyset$, so
$H_k(X^{\prime},M)=H_k(X^{\prime}) = 0$,
for $k \leq m-1$ as required. Note that one cannot generically eliminate
handles above the middle dimension through considering the dual
decomposition, because this would require
$H_{k}(X^{\prime},M^{\prime})$ to vanish.

There are also various circumstances under which one can do even better than in the
previous subsection and further eliminate the $m$th homology group through surgery, where
$n=2m$ or $n=2m+1$, in the even and odd dimensional cases
respectively \cite{milnor1,wall}. In this case the homology is
entirely determined from the boundary. This will not be discussed further here.

To illustrate the results of this section, consider now two examples.

\subsection{Example: Torus compactification from six to four dimensions}

The initial cobordism is $X=D^4 \times T^2$ with boundary $\pa X =
S^3 \times T^2$. Theorem \ref{thm:spar}, Lefschetz duality, and
the universal coefficient theorem allow us to find a cobordism
$X^{\prime}$ with the following homology known: \bea
H_0(X^{\prime}) = \bZ , \nonumber \\
H_1(X^{\prime}) = H_2(X^{\prime}) = H_5(X^{\prime}) = H_6(X^{\prime}) = 0 , \nonumber \\
H_4(X^{\prime}) = H_4(\pa X) = \bZ \oplus \bZ .
\eea
The remaining group, $H_3(X^{\prime})$, is not given by the boundary topology.
However, it cannot be zero because the exact sequence
\be
\begin{CD}
0 @>>> H_3(\pa X) @>>> H_3(X^{\prime}) @>>> H_3(X^{\prime},\pa X)
@>>> H_2(\pa X) @>>> 0 .
\end{CD}
\label{eq:nonzero}
\ee
would then imply $H_3(\pa X)=0$, which is false. Thus $H_3(X^{\prime}) \neq 0$.

The absence of first and second homology groups in the cobordism
$X^{\prime}$ is something that we could not have achieved by
taking the other obvious cobordism $X^{\prime \prime} = S^3 \times
D^2 \times S^1$ (which incidentally is also parallelisable, cf.
Corollary \ref{cor:67} below). We can see that what we have done
is trade the lower dimensional homology of $X$, $H_1(X)=\bZ\oplus
\bZ, H_2(X)=\bZ$, for the higher dimensional homology of
$X^{\prime}$.

\subsection{Example: $S^4\times T^3$ compactification from eleven to four dimensions}

Consider the eleven dimensional cobordism $X=D^5 \times S^3 \times T^3$ with boundary
$\pa X = S^4 \times S^3 \times T^3$. The cobordism is parallelisable because
$S^3  = SU(2)$ is parallelisable.

This cobordism may be considered as either a compactification from eleven to five dimensions
on $S^3 \times T^3$, or as a compactification from eleven to four dimensions on $S^4 \times T^3$.
In the latter case, the internal manifold is not a Lie group, but this doesn't matter because we
are filling in the $S^4$ to get the initial cobordism, not the
`noncompact' (i.e. large) $S^3$, which
is a Lie group.

The initial cobordism has six nonzero homology groups (not counting $H_0(X)$):
\bea
H_1(X) = H_2(X) = H_4(X) = H_5(X) = \bZ \oplus \bZ \oplus \bZ ,\nonumber \\
H_3(X) = \bZ \oplus \bZ, \nonumber \\
H_0(X) = H_6(X) = \bZ .
\eea
After applying Theorem \ref{thm:spar} we get a cobordism $X^{\prime}$
with five nonzero homology groups:
\bea
H_1(X^{\prime}) = H_2(X^{\prime}) = H_3(X^{\prime}) = H_4(X^{\prime})
= H_{10}(X^{\prime}) = H_{11}(X^{\prime}) = 0 ,\nonumber \\
H_5(X^{\prime}), H_6(X^{\prime}) \neq 0 ,\nonumber \\
H_7(X^{\prime}) = H_7(\pa X) = \bZ \oplus \bZ ,\nonumber \\
H_8(X^{\prime}) = H_8(\pa X) = \bZ \oplus \bZ \oplus \bZ ,\nonumber \\
H_9(X^{\prime}) = H_9(\pa X) = \bZ \oplus \bZ \oplus \bZ . \eea
The nonvanishing of $H_6(X^{\prime})$ and $H_5(X^{\prime})$
follows from the nonvanishing of $H_6(\pa X)$ and $H_5(\pa X)$
respectively using the exact sequence of Theorem \ref{thm:exact}
in a similar way to (\ref{eq:nonzero}).

\section{Completeness of surgery for compactification}

What are the limitations of a surgical analysis given an initial
cobordism $X$? One natural question to ask is the following:
Suppose we have two cobordisms $X$ and $X^{\prime}$ with connected
boundaries, $\pa X = \pa X^{\prime}$, can one get from $X$ to
$X^{\prime}$ via surgery?

The answer to this question uses the relation between handle
decomposition and surgery, alluded to in section 2 above. The
relation starts with the trivial cobordism $X = M\times [0,1]$,
where $M$ has dimension $n-1$. If one adds a $k$-handle to one of
the boundaries of $X$, to obtain $X^{\prime}$ with boundaries $M$
and $M^{\prime}$, then $M^{\prime}$ is obtained from $M$ via a
type $(k-1,n-k-1)$ surgery. Said differently, given a cobordism
$X^{\prime}$ between $M$ and $M^{\prime}$, the sequence of handles
in a handle decomposition of $X^{\prime}$ corresponds to a
sequence of surgeries required to go from $M$ to $M^{\prime}$.
Note that here we are using surgeries on the endpoints of the
cobordism, not on the cobordism itself as we did previously.
Theorem \ref{thm:han} then implies, as is proved directly in
\cite{milnor1,wallace1}, that:

\begin{cor} Two manifolds without boundary $M$ and $M^{\prime}$ are cobordant if
and only if $M^{\prime}$ may be obtained from $M$ via a sequence of surgeries.
\end{cor}

Thus for manifolds without boundary, the question of completeness of surgery reduces
to whether $M$ and $M^{\prime}$ are cobordant. This question is
answered in Theorem \ref{thm:cobor}.

This result may be generalised to manifolds with boundary as follows. The question
is when two manifolds with boundary, i.e. cobordisms, $X$ and $X^{\prime}$,
with $\pa X = \pa X^{\prime}$, may be obtained from one another via surgery. We will say
that $X$ and $X^{\prime}$ are {\bf cobordant with boundary} if there exists a manifold
$W$ such that $\pa W  = X \cup X^{\prime} \cup \left( \pa X \times [0,1] \right)$, where
$X$ and $X^{\prime}$ are attached to $(\pa X,0)$ and $(\pa X,1)$ in the obvious way. This
is illustrated in Figure 3, using the two cobordisms of Figure 1.
We are essentially considering a `cobordism of cobordisms'. One can generalise this
concept to the case when $\pa X \neq \pa X^{\prime}$ \cite{rourke}.
\begin{figure}[h]
\begin{center}
\vspace{4mm}
\epsfig{file=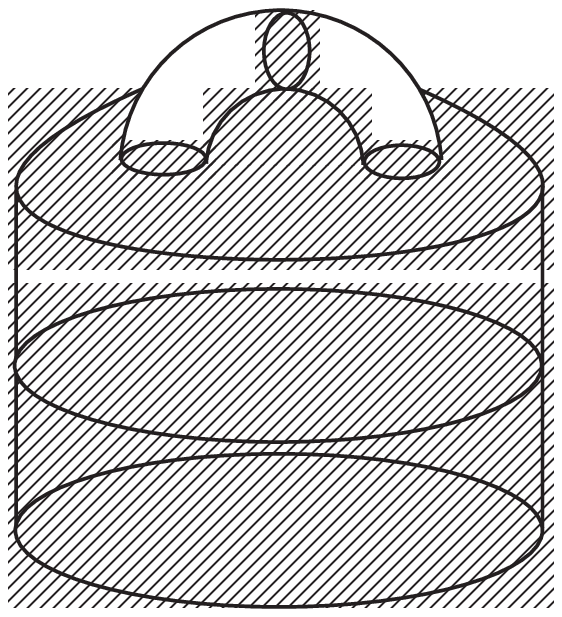,width=4cm}
\end{center}
\noindent {\bf Figure 3:} A cobordism with boundary. Here $\pa W = X
\cup X^{\prime} \cup \left( S^1 \times [0,1] \right)$.
\end{figure}

Such a cobordism with boundary now admits a {\bf handle decomposition relative to $\pa X \times [0,1]$}
in an entirely analogous way to the usual handle decomposition of a cobordism \cite{rourke}. And, as
we have just seen, this handle decomposition is equivalent to a series of surgeries. Thus one has:

\begin{thm} Two manifolds $X$ and $X^{\prime}$, with $\pa X = \pa X^{\prime}$, are related via surgeries
if and only if there exists
a manifold $W$ such that $\pa W = X \cup X^{\prime} \cup \left( \pa X \times [0,1] \right)$.
\label{thm:Q}
\end{thm}

Note that this contains within it the previous case where $\pa X = \emptyset$. This condition
is more complicated than the case without boundaries. There are a couple of cases, however, in which
we get a nice result. In dimensions 6 and 7, the oriented cobordism ring is trivial \cite{milnor3},
implying that any manifold is cobordant to the empty set. In particular, this will be true for any
$X \cup X^{\prime} \cup \left( \pa X \times [0,1] \right)$
in these dimensions. Thus Theorem \ref{thm:Q} implies:

\begin{cor} Any two cobordisms, $X, X^{\prime}$ with $\pa X = \pa X^{\prime}$,
of dimension 6 or 7 are related via surgeries.
\label{cor:67}
\end{cor}

Therefore in these dimensions at least, we are not missing any
cobordisms by restricting attention to those obtained from $D^d \times
A$ by surgery. In other dimensions, cobordisms will fall into
equivalence classes of those obtainable from one another via
surgery \cite{milnor3}.

\section{Conclusions and discussion}

This work has considered surgery as a canonical and systematic
method of rearranging cobordisms for topology change. We have
shown that any pair of connected cobordant manifolds in dimensions
$\geq 5$ admits a causally continuous cobordism. This result
extends the known result for $4$ dimensions. A consequence is that
quantum compactification is allowed in causal Lorentzian theories
of quantum gravity, without selection rules.

We further illustrated the possibilities of surgical modifications by
considering quantum compactification via parallelisable cobordisms.

Compactification is not the only higher dimensional topology
change currently of interest. Recent results concerning
instabilities in higher dimensions have hinted at the possibility
of topology change
\cite{Gregory:vy,Kol:2002xz,Wiseman:2002ti,Gibbons:2002pq,Gibbons:2002th}.
More precisely, topology change arises in these works in two,
related, ways\footnote{I'd like to thank Toby Wiseman for a
discussion on these points.}. Firstly, there seems to be a
one-parameter family of solutions connecting compactified black
strings and black hole spacetimes
\cite{Kol:2002xz,Wiseman:2002ti}. Thus along this parameter the
horizon topology, and also the singularity topology and the
Euclidean spacetime topology, changes. Secondly, one considers the
dynamical evolution of classical instabilities of uniform black
string spacetimes \cite{Gregory:vy} or generalised black hole
spacetimes in which the instability is a `ballooning' mode
\cite{Gibbons:2002pq,Gibbons:2002th}. The perturbative dynamics,
together with the lack of an obvious stable endpoint, suggest that
the system is driven towards a regime in which quantum topology
change may occur. In the black string case, this is related to the
topology change along the one-parameter family of solutions
mentioned previously. This is because the nonuniform black strings
in the family of solutions have a higher mass than the unstable
uniform black strings and the known lower mass solutions that
could be available as endpoints of the instability are black holes
and therefore have a different horizon topology.

\section*{Acknowledgements}

I would like to thank Gary Gibbons and Burt Totaro for some very
useful comments and references, and also James Sparks for some
useful information. The author is funded by the Sims scholarship.

\appendix

\section{Surgery as a canonical modification}

In this appendix we give one justification for considering surgery
as a `minimal' transformation of a manifold that eliminates the
fundamental group. Suppose we have two manifolds $X=D^d\times A$
and $X^{\prime}$, not {\it a priori} related by surgery, with the
same boundary. Further, we assume that $X^{\prime}$ is simply
connected, whilst $X$ is not.

For simplicity, assume that both $X$ and $X^{\prime}$ have no
torsion in their homology. This will enable us to assume an
isomorphism between homology and cohomology. Consider the
following pair of exact sequences, where Lefschetz duality,
$H_1(X^{\prime})=0$, and $H_1(X,\pa
X)=H_{n-1}(X)=H_{n-1}(D^d\times A)=H_{n-1}(A)=0$ have been used.
Finally, $H_1(\pa X)=H_1(\pa X^{\prime})$ is because $\pa X = \pa
X^{\prime}$.

\ben
\begin{CD}
@. \cdots  @>>> \begin{array}{c}
 H_2(X,\pa X) \\ = H_{n-2}(X) \end{array} @>>> H_1(\pa X) @>>> H_1(X) @>>> 0 \\
@. @. @. @| \\
\cdots @>>> H_2(X^{\prime}) @>>> \begin{array}{c} H_2(X^{\prime},\pa X^{\prime})
\\ = H_{n-2}(X^{\prime}) \end{array}
@>>> H_1(\pa X^{\prime}) @>>> 0 .
\end{CD}
\een

Suppose $H_{n-2}(X^{\prime}) = 0$, this will lead us to a contradiction.
Exactness of the second row would then imply that
$H_1(\pa X^{\prime})=H_1(\pa X)=0$. Now exactness of the first row implies that
$H_1(X)=0$. But this contradicts the initial assumption that the starting cobordism $X$
was not simply connected. Thus we must have $H_{n-2}(X^{\prime}) \neq 0$.

The upshot of this argument is that any process eliminating the
first homology from a manifold $X$ with $H_{n-1}(X)=0$ must
necessarily introduce an $(n-2)$-th homology, if the boundary is
preserved and modulo our torsion-free assumption. This is
precisely what a surgery of type $(1,n-2)$ does in a minimalist
way. The surgery collapses an $S^1$ and introduces an $S^{n-2}$.
This is not a uniqueness proof, but gives some intuition for why
surgery is natural in the present context. In fact, it is not
difficult to give a precise meaning to the sense in which the
change is minimal by proving that \cite{wallace1}:

\begin{lem} If a type $(k,n-k-1)$ surgery takes $X$ to $X^{\prime}$, then we have
$H_p(X) \cong H_p(X^{\prime})$ for all $p$ except $p=k,k+1,n-k-1,n-k$.
\end{lem}


\begin{thebibliography}{99}

%\cite{Tipler:qu}
\bibitem{Tipler:qu}
F.~J.~Tipler,
``Topology Change In Kaluza-Klein And Superstring Theories,''
Phys.\ Lett.\ B {\bf 165} (1985) 67.
%%CITATION = PHLTA,B165,67;%%

%\cite{Hawking:jz}
\bibitem{Hawking:jz}
S.~W.~Hawking,
``Quantum Gravity And Path Integrals,''
Phys.\ Rev.\ D {\bf 18} (1978) 1747.
%%CITATION = PHRVA,D18,1747;%%

%\cite{Vilenkin:rn}
\bibitem{Vilenkin:rn}
A.~Vilenkin,
``Approaches To Quantum Cosmology,''
Phys.\ Rev.\ D {\bf 50} (1994) 2581
[arXiv:gr-qc/9403010].
%%CITATION = GR-QC 9403010;%%

\bibitem{geroch} R.P. Geroch, ``Topology in general relativity,''
J. Math. Phys. {\bf 8(4)} (1967) 782.

%\cite{sorkin1}
\bibitem{sorkin1}
J.~Louko and R.~D.~Sorkin, ``Complex actions in two-dimensional
topology change,'' Class.\ Quant.\ Grav.\  {\bf 14} (1997) 179
[arXiv:gr-qc/9511023].
%%CITATION = GR-QC 9511023;%%

%\cite{Horowitz:qb}
\bibitem{Horowitz:qb}
G.~T.~Horowitz,
%``Topology Change In Classical And Quantum Gravity,''
Class.\ Quant.\ Grav.\  {\bf 8} (1991) 587.
%%CITATION = CQGRD,8,587;%%

%\cite{Dowker:1997hj}
\bibitem{Dowker:1997hj}
F.~Dowker and S.~Surya,
``Topology change and causal continuity,''
Phys.\ Rev.\ D {\bf 58} (1998) 124019
[arXiv:gr-qc/9711070].
%%CITATION = GR-QC 9711070;%%

%\cite{sorkin2}
\bibitem{sorkin2}
A.~Borde, H.~F.~Dowker, R.~S.~Garcia, R.~D.~Sorkin and S.~Surya,
``Causal continuity in degenerate spacetimes,''
Class.\ Quant.\ Grav.\  {\bf 16} (1999) 3457
[arXiv:gr-qc/9901063].
%%CITATION = GR-QC 9901063;%%

%\cite{Dowker:1999wu}
\bibitem{Dowker:1999wu}
H.~F.~Dowker, R.~S.~Garcia and S.~Surya,
``Morse index and causal continuity: A criterion for topology change in  quantum gravity,''
Class.\ Quant.\ Grav.\  {\bf 17} (2000) 697
[arXiv:gr-qc/9910034].
%%CITATION = GR-QC 9910034;%%

%\cite{Dowker:1997kc}
\bibitem{Dowker:1997kc}
H.~F.~Dowker and R.~S.~Garcia,
``A handlebody calculus for topology change,''
Class.\ Quant.\ Grav.\  {\bf 15} (1998) 1859
[arXiv:gr-qc/9711042].
%%CITATION = GR-QC 9711042;%%

%\cite{Ionicioiu:1997hi}
\bibitem{Ionicioiu:1997hi}
R.~Ionicioiu,
``Building blocks for topology change in 3D,''
arXiv:gr-qc/9711069.
%%CITATION = GR-QC 9711069;%%

%\cite{Alty:1994xs}
\bibitem{Alty:1994xs}
L.~J.~Alty,
``Building blocks for topology change,''
J.\ Math.\ Phys.\  {\bf 36} (1995) 3613.
%%CITATION = JMAPA,36,3613;%%

%\cite{Konstantinov:jn}
\bibitem{Konstantinov:jn}
M.~Y.~Konstantinov and V.~N.~Melnikov,
``Topological Transitions In The Theory Of Space-Time,''
Class.\ Quant.\ Grav.\  {\bf 3} (1986) 401.
%%CITATION = CQGRD,3,401;%%

%\cite{Ionicioiu:1997zy}
\bibitem{Ionicioiu:1997zy}
R.~Ionicioiu,
``Topology change from Kaluza-Klein dimensions,''
arXiv:gr-qc/9709057.
%%CITATION = GR-QC 9709057;%%

\bibitem{rotman} J.J. Rotman, {\it An introduction to algebraic topology}, Springer-Verlag, New York, 1988.

\bibitem{maunder} C.R.F. Maunder, {\it Algebraic topology}, Dover publications, New York, 1970.

\bibitem{milnor2} J. Milnor, ``A survey of cobordism theory'', Enseign. Math. {\bf 8} (1962) 16.

\bibitem{milnor3} J.W. Milnor and J.D. Stasheff, {\it Characteristic classes}, Princeton University Press, New Jersey, 1974.

\bibitem{milnor1} J. Milnor,
``A procedure for killing homotopy groups of differentiable manifolds'', Proc. Sympos. Pure
Math. Vol. III pp. 39-55, 1961.

\bibitem{wallace1} A.H. Wallace, ``Modifications and cobounding
manifolds'', Canad. J. Math. {\bf 12} (1960) 503.

\bibitem{hcobordism} J. Milnor, {\it Lectures on the h-cobordism theorem}, Princeton University Press, New Jersey, 1965.

\bibitem{rourke} C.P. Rourke and B.J. Sanderson, {\it Introduction to piecewise-linear topology}, Springer-Verlag, Berlin, 1972.

\bibitem{morse} J. Milnor, {\it Morse Theory}, Princeton University Press, New Jersey, 1963.

\bibitem{hawking} S.W. Hawking and R.K. Sachs,
``Causally continuous spacetimes,''
Commun. math. Phys. {\bf 35} (1974) 287.

\bibitem{kervaire} M.A. Kervaire and J.W. Milnor, ``Groups of homotopy spheres: I'', Ann. of Math. {\bf 77} (1963) 504.

\bibitem{wall} C.T.C. Wall, ``Killing the middle homotopy groups of odd dimensional manifolds", Trans. Amer. Math. Soc.
{\bf 103} (1962) 421.

%\cite{Gregory:vy}
\bibitem{Gregory:vy}
R.~Gregory and R.~Laflamme,
``Black Strings And P-Branes Are Unstable,''
Phys.\ Rev.\ Lett.\  {\bf 70} (1993) 2837
[arXiv:hep-th/9301052].
%%CITATION = HEP-TH 9301052;%%

%\cite{Kol:2002xz}
\bibitem{Kol:2002xz}
B.~Kol,
``Topology change in general relativity and the black-hole black-string  transition,''
arXiv:hep-th/0206220.
%%CITATION = HEP-TH 0206220;%%

%\cite{Wiseman:2002ti}
\bibitem{Wiseman:2002ti}
T.~Wiseman,
``From black strings to black holes,''
arXiv:hep-th/0211028.
%%CITATION = HEP-TH 0211028;%%

%\cite{Gibbons:2002pq}
\bibitem{Gibbons:2002pq}
G.~Gibbons and S.~A.~Hartnoll,
``A gravitational instability in higher dimensions,''
Phys.\ Rev.\ D {\bf 66} (2002) 064024
[arXiv:hep-th/0206202].
%%CITATION = HEP-TH 0206202;%%

%\cite{Gibbons:2002th}
\bibitem{Gibbons:2002th}
G.~W.~Gibbons, S.~A.~Hartnoll and C.~N.~Pope,
``Bohm and Einstein-Sasaki metrics, black holes and cosmological event horizons,''
arXiv:hep-th/0208031.
%%CITATION = HEP-TH 0208031;%%


\end{thebibliography}
\end{document}